\documentclass[12pt]{article}

\usepackage{amsmath,amssymb}
\usepackage{braket}
\usepackage{graphicx}
\usepackage{color}
\usepackage{dcolumn}
\usepackage{bm}
\usepackage[numbers,super,comma,sort&compress]{natbib}

\usepackage{authblk}
\definecolor{background-color}{gray}{0.98}

\title{Quantum Monte Carlo calculation of the bridge reaction barrier to H$_2$ dissociation on Pt(111)}

\author{Philip E Hoggan}

\affil{Institute Pascal, UMR 6602 CNRS,BP 80026, 63177 Aubiere Cedex, France}

\begin{document}

\maketitle

\begin{abstract}
The yardstick of new first-principles approaches to key points on reaction paths at metal surfaces is chemical accuracy compared to reliable experiment. By this we mean that such values as the activation barrier are required to within 1 kcal/mol. Quantum Monte Carlo (QMC) is a promising (albeit lengthy) first-principles method for this and we are now beyond the dawn of QMC benchmarks for these systems, since hydrogen dissociation on Cu(111) has been studied with quite adequate accuracy in two improving QMC studies \cite{hog0, kdd2}.

Pt and Cu require the use of pseudo-potentials in these calculations and we show that those of Pt are less problematic than those for Cu, particularly for QMC work.

In this letter, we determine the activation barrier to hydrogen dissociation on Pt(111) using the bridging geometry. This is found to agree to better than chemical accuracy with recent Specific Reaction Parameter (SRP-DFT) work, which is itself within chemical accuracy of measurement \cite{iran}.

The results suggest the dissociation barrier for hydrogen on Pt(111) is close to 5.3 (QMC) and 6.2 (SRP-DFT) kcal/mol with a QMC standard error of 1.08 kcal/mol.

This is encouraging for establishing less well-known benchmark values of industrial reaction barriers on Pt(111).
\end{abstract}

\clearpage
\makeatletter
\renewcommand\@biblabel[1]{#1.}
\makeatother
\bibliographystyle{apsrev}
\renewcommand{\baselinestretch}{1.5}
\normalsize
\clearpage

\section{\label{sec:1}Introduction}

\hskip6mm Among metal surfaces that catalyse hydrogen dissociation, Cu(111) has been the subject of very recent benchmark studies using Density Functional Theory (DFT) and state-of-the art Quantum Monte Carlo (QMC), in particular because reliable molecular beam measurements of the late barrier for this reaction are available. Both experiment and QMC are in agreement and the standard error in the QMC is close to chemical accuracy (i.e within a kcal/mol) and indications have been obtained in our earlier work for reducing the error. The case of Pt(111) is more favorable, in the sense that the element can be described by trial wave-functions with variance decreased by about an order of magnitude from the copper case, which is adversely affected by the presence of a full 3d shell.

Industrial catalysts often use Pt(111) for certain reactions triggered by bond-breaking. This surface also catalyses hydrogen dissociation. Unfortunately, Pt(111) does not yet benefit from accurate molecular beam results for this reaction. A recent DFT study \cite{iran} shows that there are several co-ordinations of the hydrogen molecule, leading to two main categories of transitions state. The 'on-top' geometry has the molecule perpendicular to the surface plane, above a Pt atom and the reaction appears to be almost barrier free, in this case. It is therefore not of an energy accessible by QMC to date, being little more than the standard error in height. This work therefore focusses on the 'bridge' orientation, in which the transition state has a hydrogen molecule parallel to the surface, above and between two Pt atoms (spanning the distance, like a bridge). The agreement is very good between the DFT study and the present QMC work, with the former needing a fitted parameter to ensure that the excesses of two functionals compensate each other and the latter, described here, being completely {\it ab initio} but somewhat time-consuming. In future, this QMC approach could be used for bench-marking key points on the potential energy surface for these systems, which would then be used to devise an optimal functional, for DFT and dynamics studies.

The QMC methods used in this letter were developed in \cite{hog0,hog1, kdd1, kdd2} where compete details can be found. They all advance solid surface QMC for catalysis involving transition metals.

The H$_2$ molecule dissociation on Pt(111) has a low experimental barrier which depends strongly on the molecular orientation and active site and ranges from some 0.06eV to 0.420eV. The higher barriers are for trigonal sites on the surface with minimal rearrangements and no defects \cite{iran}. This reaction is therefore of interest and has a low barrier. When the H$_2$ molecule lies parallel to the surface, the lowest barrier is 0.27 eV but when it is vertical, above a Pt atom (said to be 'on top') the reaction proceeds with almost no barrier. Both are therefore likely to pose a stiff challenge to QMC. The lowest QMC error bars are just over 0.043eV (which is termed chemical accuracy, i.e. error bars within 1 kcal/mol, those accessible for hydrogen dissociation on Pt(111) being just below 1.09 kcal/mol). This makes hydrogen dissociation QMC barriers reliable on Pt(111) for 'bridge' orientations but not for 'on top' positions of the H$_2$ molecule at present.

\section{\label{sec:2}Setting up the model geometries}

\hskip5mm The reaction barrier heights in this work are obtained in first-principles calculations by stringently accurate ground-state QMC calculations on two geometries comprising the same atoms. This strategy has been shown to be more accurate than referring to a 'clean' surface and the isolated molecule. \cite{alfe,aqc_phab}  The high-energy limit is the so-called transition state. The geometry is obtained by QMC optimisation from that used at the DFT level (involving the hydrogen molecule and a platinum slab cut from the face-centered cubic lattice exposing its (relaxed) close-packed (111) face. The low-energy asymptote is a geometry describing weakly physisorped hydrogen on the Pt(111) slab.

Geometric details are as follows: The parallel orientation of the H$_2$ molecule between two Pt atoms of a trigonal site is described as the 'bridge' orientation, whereas vertical the H$_2$ molecules each above a Pt atom are described as 'on top'. according to a recent DFT and dynamics study, compared with molecular beam measurements, the 'on top' orientation dissociates with practically no barrier (and would be too stringent a test for our QMC approach in its present state, since the error bars are nearly the height of the barrier). The bridge orientation is estimated to dissociate with an Specific Reaction Parameter (SRP) DFT barrier of 0.27 eV  \cite{iran}.

Our optimised bridge over a trigonal site leads to a QMC barrier of 0.23 +/- 0.047 eV. This corresponds well to the 'bridge' DFT and molecular beam barrier, that value of 0.27eV being within the error bar and is not at the lower limit of complete the range, which can be ascribed to more reactive 'on top' catalysed barriers. It is close to the bridge measured barrier and therefore consistent with rearranged Pt(111) surface barriers. that benchmarks the accurately measured and significantly higher barrier for hydrogen dissociation on Cu(111) to within some 1.5 kcal/mol \cite{kdd2}. The barrier on Cu(111) has been given by accurate molecular beam experiments at 0.63 eV or 14.5 kcal/mol. For Pt(111) we get a DMC value of 5.3 kcal/mol or 0.23 eV. Error bars are a bit lower (less than 1.085 kcal/mol).

Slab thickness: a study in \cite{kdd2} with experimental skin atom-layer spacing converging to bulk for four layers, minimum, tends to show a 5-layer slab is significantly better. Further increases do not give much improvement, we thus chose the 5-layer slab model in this work.

The five-layer slab with four hydrogen molecules adsorbed upon it was compared in surface size. First, a 2 by 2 mesh surface followed by a 4 by 4 mesh surface. Comparing these two sizes of surface allows you to extrapolate to the infinite surface extent. Note that for the cited work on copper, a significant overall reduction in barrier height was observed, whereas for Pt(111) the overall effect is very small. This suggests that finite size effects are lower for Pt(111) than Cu(111) as confirmed by the relative sizes of the single-particle contributions that are approximately an order of magnitude lower for Pt(111) than for Cu(111).

Errors are statistical and also systematic. The wave-function used to initialise the QMC calculation is based on a single slater determinant, in view of the system size. Nevertheless, some methods use embedding of a molecular portion of the active site and these may have their geometry optimised within QMC \cite{hog1}. They could also serve as the basis for a selected CI (CIPSI). To obtain a manageable number of valence electrons for heavy elements like Platinum (and copper) a suitable pseudo-potential must be used. For Pt, we use a Z=60 pseudo-potential, leaving 18 valence electrons/atom \cite{hog1}. Both the limitation of the trial wave-function to a ground-state single determinant and the use of pseudo-potentials introduce deficiencies in its nodal structure. Since we need the fixed node approximation for fermionic symmetry in  DMC, i.e. we distribute configurations according to a density formed as the product of the trial wave-function and its update, then this is a source of error that is projected out by using a complex Jastrow factor to optimise the trial wave-function in the VMC stage. 

This Jastrow factor is a set of polynomials in the inter-particle variables (see subsection below).

\section{\label{sec:3} Quantum Monte Carlo Methods}

Two QMC methods are used which are detailed elsewhere. Briefly, the first, so-called Variational Monte Carlo (VMC) minimises an observable, typically wave-function variance or total energy. The trial wave-function can be made very flexible by multiplying its real and imaginary parts with distinct Jastrow factors. These polynomials in the two-particle and also three-particle(electron pair and nucleus) separation variables may be taken to order 6-9 and coefficients treated as variational parameters. The energy minimisation is used here for fine-tuning following an initial variance minimisation. Cut-off (range) parameters can be fixed after the variance minimisation.

The accurate QMC method follows. It is Diffusion Monte Carlo, where a propagator method is used to solve the time-dependent Schroedinger equation of the system, cast as a diffusion-drift equation in imaginary time, which is then increased in small steps like a time variable. The step-size ($\tau$) chosen is 0.005 au and was checked for possible bias by doubling and halving the value and comparing three equivalent runs for $\tau, \tau/2, 2 \tau $. The energies fell in a linear domain with average differences which extrapolate to zero time-step falling within less than 1 meV, which enables us to conclude that the chosen step-size will not lead to time-step bias.

The number of data-points collected is 52 000 for each geometry.

We compare the bridging TS geometry, directly optimised within VMC to an asymptote with the hydrogen molecules 8 \AA \hskip4mm from the surface. The clean metal surface has been shown to be a poor reference, due to the surface states and the fact that physisorped atoms and molecules have a symmetry-breaking role on these states in our previous work AQC.

Solid state QMC is currently coming of age and a number of metal-surface with interacting molecule systems have been studies, since the activation barrier published by Pozzo and Alf\'e in 2008, for the H$_2$ molecule dissociating on Mg(0001) \cite{alfe}. This system does not yet have accurate (molecular beam experiment) measurements.

Applications to catalysis and many electronic devices require surfaces of transition metals. For (almost) all systems of interest, all-electron calculations are prohibitive in terms of cpu time. Pseudo-potentials are used. With this in mind, Doblhoff Deir meticulously evaluated errors due to wave-function and pseudo-potential (in particular non-locality, often neglected but significant when 3d electrons are studied) for transition metal binary compounds, involving first transition series elements. (see \cite{kdd1}). All these errors point to strategy for the improvement of metal slab design. (see \cite{kdd2}). Non-locality error and errors compounding the fixed-node error are the most difficult to circumvent. The work on transition metal binary compounds and the subsequent benchmark for hydrogen dissociation on Cu(111) \cite{kdd1,kdd2} use Casula T-moves. This is a modified DMC algorithm that guarantees the variation principle.

In this work, we investigated whether the DMC needs to be variational. Otherwise a quadratic convergence to the ground-state is observed, occasionally from below. This did not appear to pose any problem for the current DMC work.

From preliminary studies of platinum surfaces for adsorped molecules by the author, it appears from trial wave-function variance data that the problems encountered with elements involving partly or totally full 3d shells are less acute for Pt.

\subsection{\label{sec:3.1} Optimising the Jastrow factor.}

As already mentioned, this is done during the VMC step.

 The Jastrow factor contains two-particle (electron-electron and electron-nucleus) terms as well as all important three-particle electron-electron-nucleus terms. In fact, for the slab representing the solid+H$_2$, there is no inversion symmetry and the wave-function therefore has real and imaginary terms. Since this complex arithmetic is necessary, it is simple and not too costly to optimise the real contribution with a Jastrow factor and the imaginary contribution with an independent Jastrow factor, thus shifting the wave-function nodes for the product as a whole towards those of the exact system.
We set up a Garnet Chan complex Jastrow factor. This is well-documented to project the correct nodes in certain limits. Implementation is easy since the absence of inversion symmetry will lead to complex wave-functions such that the real and imaginary parts may be expanded on a basis with independent Jastrow factors, thus using a complex Jastrow factor with no extra programming. Loss of speed, due to complex arithmetic means the steps are approximately four times longer for a given expansion order in the polynomials constituting the Jastrow factors. VMC energy minimisation driven optimisation of complex Jastrow factor:

Comprises two real parameter sets (for the real and imaginary components). The spin states are all consdered.

u-term  (nuclear-electron instantaneous distance as variable) to order 6.

$ \chi $-term (electron-electron instantaneous distance as variable, free spin) to order 8.

F-electron pair-nucleus 3 particle term, set up as a nested loop over nuclei and electron separation distance) to order 5.

\subsection{Finite-size effects}

Single-particle finite-size effect corrections are accounted for by twist averaging. We use the same strategy as developed in \cite{hog0}.

This procedure will not be detailed in the present letter and accounts for a minor effect of about 1\% on the barrier. The additional corrections are described in \cite{kdd2} and are a factor 4 smaller or less.

\section{\label{sec:4} Perspectives and Conclusions }

This Quantum Monte Carlo simulation of the bridge geometry hydrogen dissociation barrier on Pt(111) has standard error within a few percent of chemical accuracy (1.08 kcal/mol) and agrees with the Specific Reaction Parameter DFT work 8cite{iran} which claims chemical accuracy to less than 1 kcal/mol, with the DFT barrier at 6.2 and the QMC barrier at 5.4 kcal/mol. Note that the QMC barrier for hydrogen dissociation in \cite{kdd2} also falls bellow the accurate molecular beam measurements for it.

The quality of trial wave-functions in this work show the advantage of projecting towards exact nodes using the complex Jastrow fctor developed in \cite{chan}. From variance values, which are must lower for Pt(111) than for Cu(111) we conclude that much of the pseudo-potential locality issue is specific to 3d electrons.

Bond dissociation at metal catalyst surfaces is often involved in industrial synthesis of chemicals. This work describes a prototype for future QMC benchmarks on the Pt(111) industrial catalysts and synthesis reactions.

%\section{\label{sec:5}References}

\end{document}